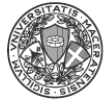

**DEPARTMENT OF HUMANITIES**

MASTER OF ARTS

IN

ACCESSIBILITY TO MEDIA, ARTS AND CULTURE

# Accessibility Literacy:
Increasing accessibility awareness
among young content creators

Alina Karakanta

Supervisor:
Prof. Gian Maria Greco

ACADEMIC YEAR 2022/2023

# Table of Contents

## Abstract – English

Smartphones, social media and mobile applications have turned users into web content creators, the so-called *prosumers*. With the abundance of tools available, individuals without any technical skills can easily become proficient in creating and sharing complex content featuring not only text, but also video, sound and images. However, proficiency in creating web content does not imply proficiency in creating accessible content. User-generated content is often not accessible for two reasons. Firstly, creators are not aware of the fact that some audiences cannot access this content due to sensory, cognitive, linguistic or cultural barriers. Second, they do not have the necessary knowledge and skills for rendering it accessible. This restricts access to user-generated content for several groups of people.

This thesis aims at exploring and increasing the literacy level of content creators regarding media accessibility, focusing especially on web accessibility. The goal is to increase the awareness of users about the accessibility barriers that can be found in web content and educate them on practices for creating accessible content. For this purpose, simple and easy-to-use training material was designed, including infographics and short quizzes on three basic tools in media accessibility: alternative text, plain language and closed captioning. All training material is designed so as to be both pleasant and accessible. The infographics are suitable for university education but can also be disseminated through social media and other channels.

A survey was designed to investigate the accessibility literacy level of young creators before and after receiving training on accessibility. In addition, the questionnaire explored the willingness of creators to change the mode they create content to make it more accessible and in which ways. The findings suggest that young content creators have limited accessibility literacy but even simple training material can positively influence their perceptions on web accessibility. Participants seem to be willing to implement accessibility tools for their UGC, and the ways vary depending on the type of content they create and share and its purposes. Despite traces of the medical model of disability and a particularlist view of accessibility in some responses, accessibility is considered important for increasing inclusion, improving UGC, and shaping a fairer society for everyone. All in all, the outcomes of the project aim at increasing the knowledge and critical thinking of young creators around media accessibility with the goal of improving access to user-generated content.

## Abstract – Italiano

Smartphone, social media e applicazioni mobili hanno trasformato gli utenti in creatori di contenuti web, i cosiddetti *prosumer*. Con l'abbondanza di strumenti disponibili, gli individui senza competenze tecniche possono facilmente diventare esperti nella creazione e condivisione di contenuti complessi, che comprendono non solo testo, ma anche video, suoni e immagini. Tuttavia, la competenza nella creazione di contenuti web non implica la competenza nella creazione di contenuti accessibili. I contenuti generati dagli utenti spesso non sono accessibili per due motivi. In primo luogo, i creatori non sono consapevoli del fatto che alcuni gruppi di persone non possono accedere a questi contenuti a causa di barriere sensoriali, cognitive, linguistiche o culturali. In secondo luogo, non hanno le conoscenze e le competenze necessarie per renderlo accessibile. Ciò limita l'accesso ai contenuti generati dagli utenti per diversi gruppi di persone. Questa tesi mira ad esplorare e aumentare il livello di alfabetizzazione dei creatori di contenuti riguardo all'accessibilità dei media, concentrandosi in particolare sull'accessibilità del web. L'obiettivo è aumentare la consapevolezza degli utenti sulle barriere d'accessibilità che si possono trovare nei contenuti web ed educarli sulle pratiche per la creazione di contenuti accessibili. A questo scopo è stato progettato materiale formativo semplice e di facile utilizzo, tra cui infografiche e brevi quiz su tre strumenti fondamentali nell'accessibilità dei media: testo alternativo, linguaggio semplice e sottotitoli. Tutto il materiale formativo è progettato in modo da essere piacevole e accessibile. Le infografiche sono adatte alla formazione universitaria ma possono essere diffuse anche attraverso i social media e altri canali. È stato progettato un sondaggio per indagare il livello di alfabetizzazione in materia di accessibilità dei giovani creatori prima e dopo aver ricevuto la formazione sull'accessibilità. Inoltre, il questionario ha esplorato la volontà dei creatori di cambiare la modalità con cui creano i contenuti per renderli più accessibili e in quali modi. I risultati suggeriscono che i giovani creatori di contenuti hanno un'alfabetizzazione limitata sull'accessibilità, ma anche il semplice materiale formativo può influenzare positivamente la loro percezione dell'accessibilità web. I partecipanti sembrano essere disposti a implementare strumenti di accessibilità per i propri contenuti e le modalità variano a seconda del tipo di contenuto che creano e condividono. Nonostante le tracce del modello medico della disabilità e una visione particolaristica dell'accessibilità in alcune risposte, l'accessibilità è considerata importante per aumentare l'inclusione, migliorare i contenuti e plasmare una società più giusta per tutti. Nel complesso, i risultati del progetto mirano ad aumentare la conoscenza e il pensiero critico dei giovani creatori sull'accessibilità dei media con l'obiettivo di migliorare l'accesso ai contenuti generati dagli utenti.

# 1. Introduction

In the span of just a few decades, the world has undergone a profound digital transformation. In an era defined by rapid technological advancements and an insatiable appetite for information, web content has emerged as a dominant force shaping the way we perceive, learn, and engage with the world around us. The fusion of text, image and sound in the digital realm has transcended traditional modes of communication, giving rise to new complex languages. From streaming platforms that deliver an endless array of entertainment to the democratisation of video creation, editing, and sharing, the influence of audiovisual content is pervasive and transformative. Whilst these novel media languages present opportunities to transcend conventional barriers (e.g. cultural and linguistic), they might also establish new barriers or reinforce existing ones.

The democratisation of content creation has led to the proliferation of user-generated content (UGC), which has transformed the way people interact with online platforms and consume information. User-generated content is any type of content (data, information, media) created by regular people based on their experiences, opinions, ideas, or feedback, and shared on the web, especially on digital platforms and social media networks (Krumm et al. 2008). UGC is often more authentic and relatable than professionally produced content, it encourages creativity and innovation and allows individuals to express their unique identities and tell their personal stories. For these reasons, UGC can reach wide audiences and provide a platform for underrepresented voices and communities to share their opinions and perspectives. The popularity of UGC is driven by its ability to capture genuine human experiences, foster connections, accommodate diverse interests, and enable active participation in the online world. It has revolutionised how information is produced, shared, and consumed, making it a cornerstone of the modern digital landscape.

At the heart of this shift lies the phenomenon of social media, an intricate web of platforms and applications that supports the diffusion of user-generated content. According to Kaplan and Haenlein (2010), social media can be defined as 'a group of internet-based applications that build on the ideological and technological foundations of Web 2.0, and that allow the creation and exchange of User-generated content'. Social media have become an indispensable part in the lives of many, especially the younger generations. In a 2021 survey, 84% of American adults ages 18 to 29 say they use at least one type of social media. Among the most popular platforms are YouTube used by 95% of the respondents, Instagram (71%) and Snapchat (65%), while roughly half of the respondents say they use TikTok.[1] The fact that the most popular social media are video platforms shows the familiarity of young users with consuming, creating and sharing web content combining visual and acoustic modalities.

---

[1] https://www.pewresearch.org/internet/2021/04/07/social-media-use-in-2021/

The technical and financial ease to access and use technology and platforms has lowered the barriers to entry for content creation. With just a smartphone, an internet connection and a social media account, individuals can easily produce and share content, facilitating the process of information dissemination. Back are the days when complex technical skills and specialised equipment made content creation possible for the few savvy users. Diversity, inclusion, and active participation have become integral characteristics of a bottom-up creation process. Along the principles of empowerment, participation, creativity, community building, a new generation of content creators is being bred.

While the technological gap between professional and novice has been closed when it comes to web content creation, the accessibility gap in the consumption of this content remains a prevailing issue. According to estimates of the World Health Organisation (2011) about 15 percent of the world's population is affected by a form of disability. While UGC has many benefits and has become a significant part of online interactions, it is important to acknowledge that it is not always accessible to people with disabilities. Many content creators may not possess the basic technical knowledge to make their content accessible or even be aware of the importance of creating accessible content. This can result in content that produces various barriers, such as visual, hearing, cognitive. For example, images and videos are often a core component of UGC, but they can be inaccessible to individuals with visual impairments. Without proper alt text (descriptive text) of images or captions for videos, screen readers used by visually impaired users cannot convey the content's meaning effectively. Videos or audio content without accurate transcriptions or subtitles can exclude people who are deaf or hard of hearing from accessing the information. Additionally, UGC platforms sometimes allow for unstructured formatting, leading to poorly organised content. Lack of a clear structure makes content difficult to follow by people with cognitive disabilities while screen readers will fail to convey the information in a coherent manner to users with visual disabilities. UGC can sometimes use complex language, slang, or jargon that might be challenging for individuals with cognitive disabilities or those who are learning the language. Addressing these accessibility challenges requires a combination of efforts from content creators, platform developers, and the wider online community.

Since the 2000s, the social model of disability (Oliver, 1990) has been gaining strength. The social model of disability has been developed inside the disability movement as a response to the negative aspects of the dominant medical model of disability, which considers disability as a problem of the individual, who has to be adjusted usually through medical interventions to become 'normal'. The social model, instead, views disability as a social construct. It is the society that disables people with impairments[2], by imposing a social environment which has been developed by an ableist perspective of the world. Consequently, within the social model, the focus is on making changes to the environment to ensure accessibility to as many people as possible (McGuire, 2011). This includes the adoption of universal design concepts, which involve responding to the diversity of users from the design phase. For UGC, a proactive and universalist approach would require, as a first step, creating critical learning spaces (Greco, 2019) for raising awareness about media accessibility and encouraging content creators to consider and prioritise accessibility.

The study presented in this thesis aims at increasing the accessibility literacy of young content creators with the overarching goal of more accessible user-generated content. This is implemented through training which raises the awareness of creators on accessibility tools and equips them with basic knowledge to make user-generated content accessible. More specifically, the research questions investigated in this thesis are the following:

1. What is the current literacy level on media accessibility among young content creators?
2. Can training material, such as infographics and quizzes, be used to educate young content creators on accessibility?
3. Is training helpful to improve perceptions of young content creators towards web accessibility?
4. Are young content creators more willing to create accessible content after receiving training on web accessibility?

The research questions were pursued in two steps. First, training material was developed on basic web accessibility tools. This includes infographics and quizzes. Infographics are visual representations of information that use a combination of text, images, and design elements to convey complex concepts in a clear and concise manner. In the context of education, infographics enhance learning experiences by simplifying complex concepts and make educational content more engaging and accessible. Quizzes allow users to check the knowledge they obtained from the infographics and interact with practical examples of applying the concepts for their content. Second, a survey was designed to collect details about the current accessibility literacy of young creators and the effectiveness of the training material in shifting their perspectives on media accessibility.

---

[2] "People with impairments" is here a neutral linguistic way to refer to people. That is, the term is not value-ladden.

This thesis is structured as follows: Section 2 lays out the background around media accessibility and accessibility training. The training material developed and the methodology for conducting the study are presented in Section 3. The results of the study are presented and discussed in Section 4, while the conclusions are drawn in Section 5.

# 2. Background
## 2.1. Media and web accessibility

The scope of this thesis is the accessibility of user-generated content (UGC) shared on the web, especially on social media platforms. Since the web is an audiovisual medium, we position the accessibility of UGC under the framework of media accessibility (MA). MA concerns access to media and non-media objects, services and environments through media solutions, for any person who cannot or would not be able to, either partially or completely, access them in their original form (Greco, 2018). MA comprises three categories: solutions that allow access to media objects, services, and environments; solutions that allow access to media objects, services, and environments through media tools; and solutions that allow access to non-media objects, services, and environments through media instruments (Greco, 2019). UGC is related to solutions in the second category, that is, media objects becoming accessible through media tools.

When it comes to audiovisual content, that is content including audio, video and images, MA has been long considered a sub-field of audiovisual translation (AVT). A particularlist account restricted MA as relevant only for persons experiencing barriers, where the distinction between AVT and MA was based on whether they address linguistic or sensory barriers. For example, Pérez-González (2014) proposed dividing AVT types into three main groups: subtitling, revoicing, and assistive forms of AVT. Matamala (2021) took a more integrated approach where assistive forms do not form a category of their own, by proposing a categorisation into sound modalities and visual modalities. Sound modalities are the ones where the product is delivered in spoken form (dubbing, voice over, interpreting, audio description, audio subtitling), while visual modalities include subtitling and sign language interpreting. However, over the past years with the broader acceptance of universalist approaches, MA has moved under the umbrella of accessibility studies and embraces forms unrelated to translation (Greco, 2018). Using translation as a classification criterion, MA modalities and services can be classified as translation-based and non-translation-based. Translation-based include sign language interpreting, subtitling and subtitling for the D/deaf and the hard-of hearing, audio description and audio narration, transcripts, dubbing and voice over. Non-translation-based modalities are audio introductions, audio subtitles, clean audio, slow reproduction, screen reading and tactile reproductions (Greco and Jankowska, 2020; Jankowska, 2020).

The past few years have witnessed a shift from particularlist to universalist, from maker-centred to user-centred and from reactive to proactive approaches in accessibility (Greco, 2018). As mentioned above, universalist perspective implies that accessibility concerns all human beings and not just a particular group, usually the people with disabilities. As such, in the scope of this thesis, accessibility concerns anyone interacting with or sharing UGC. According to the user-centred approach, each agent (maker, expert, user) plays a role in constructing the experience. This thesis takes a first step towards shifting the agency away from the accessibility expert and aims at bridging the expert-maker gap, by training UGC makers on how to use media tools themselves to provide accessibility for the content they create. The proactive shift means that access concerns are not left to be addressed after the realisation of the media artifact. For UGC, given that accessibility may not be a wide-spread concern among non-expert users yet, a proactive approach entails as a first step prioritizing the user's needs and considering accessibility from inception rather than as an afterthought, even though the solutions could still be ex-post with specific adaptations or additions to the content.

Narrowing down the scope from all types of media to the web, web accessibility is concerned with ensuring that there are no barriers preventing the interaction or people with the web no matter their needs, wishes, specific features or how they access the web. The foundations of web accessibility are established through internationally recognised standards and guidelines. The World Wide Web Consortium (W3C) founded the Web Accessibility Initiative (WAI) in 1996 to promote web accessibility by creating tools, guidelines, and reporting systems. The most widely known guidelines are the Web Content Accessibility Guidelines (WCAG). The initial WCAG version 1.0 was developed in 1999, while the most recent version is WCAG 3.[3] WCAG are a set of globally recognised standards which provide a comprehensive framework for creating accessible web content. They aim to ensure that websites, applications, and digital content can be perceivable, operable, understandable, and robust for all users, regardless of their abilities or disabilities. WCAG includes guidelines and recommendations to enhance the accessibility of web content, covering aspects such as text alternatives for non-text content, the use of captions, keyboard navigability, and more. By adhering to WCAG, creators can make their digital content more inclusive and usable for a diverse audience.

Another set of guidelines promoted by WAI are the Authoring Tool Accessibility Guidelines (ATAG).[4] Authoring tools are software and services that "authors" use to produce web content. ATAG have two goals; first, to make the authoring tools themselves accessible, so that people with disabilities can create web content, and second, to help authors create more accessible web content. The scope includes tools used by web developers, designers, writers but it extends to websites that let users add content, such as blogs, wikis, photo sharing sites, online forums, and social networking sites. The WCAG and ATAG guidelines provide the starting point for the training material created in this thesis.

---

[3] https://www.w3.org/TR/wcag-3.0/#
[4] https://www.w3.org/WAI/standards-guidelines/atag/

### 2.2. Social media accessibility

The proliferation of social media platforms necessitates a focus on accessibility within these digital spaces. Advancements in assistive technologies have significantly impacted the way individuals with disabilities interact with digital media. Screen readers, magnifiers, voice recognition software, and other assistive tools facilitate web access for people with visual, auditory, motor, or cognitive impairments. Still, for these technologies to function in an optimal way, there are two preconditions: first, the content should be created bearing in mind accessibility guidelines; second, the right accessibility functionalities need to be implemented in the platforms.

There is a lack of scientific literature exploring the accessibility features of popular social media platforms and communication tools. Most social media platforms and apps have a commercial nature and therefore information on their accessibility features can mostly be found on the companies' blogs and business communications. In addition, social media accessibility is not technically required under the WCAG 2.1 standards and platforms cannot be changed. Social media platforms have often received criticism for addressing accessibility only as an afterthought or for implementing 'early versions' of accessibility features which in fact do not ensure access. A case that stands out has been Twitter, which was heavily criticised for lacking accessibility features for the new functionalities they implemented, for example no captions for voice tweets. Later it became known that Twitter did not have a dedicated accessibility team until 2020.[5] Despite this case, it appears that attempts are being made to implement more and more accessibility features along with improved functionalities in social media platforms, either due to public pressure or to reach wider audiences. The next paragraphs present some of the accessibility features that popular social media platforms offer at the time of writing.

Descriptive text, more commonly known as "Alt text", and image descriptions provide visual support for visual items such as images and videos. It is text that describes the content or function of a visual item, which can be magnified, resized or read out by screen readers. Alt text is supported in most social media platforms, including Facebook, Instagram, Twitter, LinkedIn, and recently in

---

[5] https://www.theverge.com/2020/6/18/21296032/twitter-audio-tweets-accessibility-volunteers

Pinterest. Unfortunately, adding Alt text in Pinterest is possible only to static pins (images) while adding alt text to video pins requires a business account. Some platforms have introduced a functionality that automatically generates Alt text, which can be then edited by the user.

Structured content can be useful for screen readers but also for people with cognitive impairments or for any person who has difficulties following the presentation of content in social media. Even though most platforms have effects for posts, such as backgrounds, media, gifs, emojis, hashtags, structuring the content is not always easy. For example, it is normally not possible to structure the text using HTML. Another issue for accessibility is hashtags and emojis. Hashtags are read by screen readers as one word. They can be better read if Camel/Pascal case is used, but there is no message or prompt to inform users on how to better structure their hashtags in the platforms. Screen readers also read out the names of emojis, making it confusing for users to understand the message. So far, the platforms offer no possibility to mark emojis as decorative so that they are skipped by screen readers.

To date, all reviewed platforms allow captions to be added to videos. Users can upload subtitle files (mainly in .srt format) that they have created using external tools. Another popular feature is the automatic captioning of videos. Since creating professional captions requires technical skills and often costs that cannot be sustained by all users, automatic captioning is considered as a way to ensure accessibility of video content. Automatic captioning is offered in Facebook, Twitter, Instagram, Linked and YouTube. After the automatic captions are generated, users can edit them to correct mistakes and improve their readability. YouTube also offers live captions for streaming events, a feature often implemented in meeting platforms such as Zoom. However, automatic captioning does not come without shortcomings. The availability of languages supported is often limited. LinkedIn offers auto-captions only for English, Instagram for 17 languages, which means the feature is not available to everyone. In addition, the quality of the auto-captions depends heavily on a variety of factors, such as type of content, noise environment, slang, accent, video quality etc. Especially UGC is a challenging type of content for automatic systems. Fortunately, most platforms have a message warning users that the quality of the auto-captions may be poor and that it is possible that auto-captions contain offensive or harmful words. Another limitation is that the availability of auto-captions depended on whether the author enabled them or not. For example, YouTube captions 'are available on videos where the owner has added them, and on some videos where YouTube automatically adds them'.[6] In many cases, deaf and hard-of-hearing persons had to remind or urge creators to enable captions.[7] Responding to the need of users to have captions whenever they need them, Tik Tok enabled viewers to turn on automatic captions of the content they watch. Still, the language coverage and quality of automatic captions is not sufficient to cater for the accessibility needs of a large number of people, but the availability of tools to help creators caption their content is definitely a step in the right direction.

---

[6] https://support.google.com/youtube/answer/100078?hl=en&co=GENIE.Platform%3DDesktop

When it comes to legislation, social media falls under audiovisual media. At the European level, significant steps have been made to improve the accessibility of audiovisual services and products. The revision of the Audiovisual Media Services Directive (AVMSD)[8] extends certain audiovisual rules to video-sharing platforms and social media services, however, the relationship with accessibility is not clearly started. The European Accessibility Act[9] includes under its scope audiovisual media, without specific mentions of social media.

A few governmental projects can be mentioned regarding social media accessibility. The VIVID:T Project from the European Disability Forum aimed at creating a community to work towards disability inclusion in humanitarian action and volunteering. A social media accessibility toolkit was devised as part of the project in 2021.[10] This toolkit contains tips in plain language on how to make social media posts accessible, covering major aspects of accessibility such as fonts and colours, Alt text, emojis, GIFs, hashtags, and captions. The US federal government, in collaboration with other agencies, developed a social media accessibility toolkit for 'Improving the Accessibility of Social Media for Public Service'. The toolkit contains guides to help social media content managers and other communication specialists ensure that their posts are reaching the largest audience, including people with disabilities. The guidelines provide tips for writing in plain language, and for making accessible posts on Facebook, Twitter, Snapchat, Instagram, YouTube, and blogs.[11] The Australian Communications Consumer Action Network (ACCAN), published a report called "SociAbility: social media for people with a disability" in 2012, however, the information does not seem to be updated after 2017 to account for the constant changes in the platform functionalities.[12] All in all, such projects can lead the way for highlighting best practices and areas for improvement to ensure that individuals with disabilities can engage fully in online social interactions.

---

[7] https://www.tiktok.com/@mikaelachavezt/video/7074624459499588910?_r=1&_t=8UAgh8hmc6o Last accessed 02/10/2023
[8] https://ec.europa.eu/digital-single-market/en/revision-audiovisual-media-services-directive-avmsd
[9] https://ec.europa.eu/social/main.jsp?catId=1202
[10] https://www.edf-feph.org/content/uploads/2021/07/3.2-Toolkit-Accessible-Social-Media.pdf
[11] https://digital.gov/resources/federal-social-media-accessibility-toolkit-hackpad/
[12] https://mediaaccess.org.au/web/social-media-for-people-with-a-disability

## 2.3. Accessibility education

Education and training play an important role on the perceptions around accessibility. Accessibility is taught and studied in various academic fields, from human-computer interaction, information technology, design, engineering and architecture to psychology, cognitive science, linguistics, health sciences and law. Web accessibility is taught within the subject area of Human-Computer Interaction and related computer science and engineering disciplines, mainly as a standalone course or module (Lazar 2002; Rosmaita 2006; Bohman 2012). However, there have been approaches arguing in favour of embedding accessibility topics throughout the computer science curricula, combining dedicated courses, industry lectures, and project-based learning (Waller et al. 2009). Indeed, the importance of understanding the diversity of users through multiple learning experiences, in line with the critical learning spaces (Greco, 2019), has been highlighted in several university programmes. Putnam et al. (2016) conducted a study on best practices for teaching accessibility in university-level programs in the US, focusing on accessibility in Information and Communication Technologies (ICTs) programmes. They found that instructors emphasized the need for students to develop awareness and understanding for a diversity of ICT users through multiple different experiences, among which participatory approaches directly involving users with disabilities in the research projects, guest speakers, field trips, simulating disabilities, and the use of videos/movies.

With the emerging student-centred pedagogical paradigm in higher education, alternatives to traditional classroom modules on accessibility have been explored. One of these alternative approaches is game-based learning. Games have been shown to be feasible and effective learning spaces, helping students develop conceptual understanding, transversal knowledge, and action-directed learning (Vlachopoulos and Makri, 2017). Kletenik and Adler (2022) created three games, simulating disabilities, to engage students to learn about accessibility. In their study involving 113 beginner computer science and non-computer science students they found that playing games increased student empathy towards people with disabilities and motivated them towards accessible design. In a subsequent study, Kletenik and Adler (2023) compared traditional classroom modules and accessibility simulation games for software developers. They found that accessibility lectures and readings were effective at inspiring student empathy towards people with disabilities, but with short-lived effects. Contrary, the simulations inspired larger and longer-lasting empathy and consideration of people with disabilities. Focusing on participatory approaches, Souza et al. (2022) adopted a Participatory Action-Research design involving people with intellectual disabilities to promote inclusive and accessibility-driven game design and development skills for new game designers.

As for media accessibility, MA experts were initially trained internally by institutions and organisations, and training was tailored to the needs of the respective organisation (Cerezo Merchán, 2019). For example, respeakers often receive training in the television channels where they are employed. Later, MA was incorporated in translation studies programmes, mainly as 'a new field within the discipline of (Audiovisual) Translation Studies' (Mazur and Vercauteren, 2019). In such programmes, MA is either a self-contained module or part of larger modules in AVT. This particularlist account restricted MA to services related to AVT, mainly subtitling for the Deaf and hard of hearing and audio description, or also sign language interpreting in the case of interpreting programmes. The main profiles that have risen from such accounts are the audio describer and the (live) subtitler. In general, training for these profiles is based on prescriptive guidelines, and includes linguistic and technical skills, and the use of specialised software. Moreover, the skills and competences experts are defined in two standards; ISO/IEC TS 20071-21:2015 on audio description and ISO/IEC TS 20071-25:2017 on subtitling. In addition to university education, AVT scholars have been involved in professional training courses on accessibility. Such projects are the Easy Access for Social Inclusion Training (EASIT)[13] focusing on training for making information easy to understand, Interlingual Live Subtitling for Access (ILSA)[14] for training live subtitlers and ADLAB Pro[15] for training audio describers. Greco (2020) has been arguing for a new critical pedagogical approach to the development of accessibility-driven skills in future media creators, which is diversity-based, user-led, proactive-oriented, creativity-driven, and quality-centred.

Contrary to previous work on accessibility training, this project aims at improving the literacy of lay users on accessibility issues. As such, this project falls into the realm of non-professional education. For this reason, the term 'literacy' is preferred over traditional, university curriculum terms such as 'skills and competences'. According to the Cambridge Dictionary, literacy is defined as 'a basic skill or knowledge of a subject'. 'Basic' means that users do not become experts but are aware of web accessibility and have the bare minimum skills which allow them to implement basic accessibility tools when creating UGC. As a first step, this thesis aims at raising user's awareness, defined as 'knowledge that something exists, or understanding of a situation or subject at the present time based on information or experience'. The training material is designed to improve the knowledge, understanding and provide basic skills. It is not exclusively destinated to be integrated into university curriculum but can be disseminated through social media and other media. This focus on the lay users and improving their accessibility literacy is what mainly differentiates the present study from previous work on accessibility training.

---

[13] https://webs.uab.cat/easit/en/
[14] http://ka2-ilsa.webs.uvigo.es/project/
[15] http://www.adlabproject.eu/home/

# 3. Methodology
## 3.1. Training material

The first part of the thesis is dedicated to the training material aimed at improving the accessibility literacy of web content creators. The material includes infographics on specific accessibility tools, followed by quizzes. The main source for the content in the infographics is the WCAG guidelines, modified based on our research on the functionalities of the social media platforms. All created material can be found in the public repository https://github.com/fatalinha/accessibility-literacy.

### Infographics

Four infographics were created as training material for this study. Given the limited scope of this thesis and to avoid an extreme load for the participants in the study, a selection of accessibility topics had to be made. The first criterion was the ease of implementation. The accessibility solutions had to be easy to explain to people without previous knowledge in accessibility and simple to implement without expert knowledge of software or other skills. For instance, audio description requires more explanation and practical training than Alt text, which would make the infographic too complex to design and to process. Closed captioning, while relatively specialised when it comes to conformity to subtitling constraints, is more accessible due to the abundance of basic software that allows average users to create captions, either from scratch or by editing automatically generated ones. It is expected that users of social media and video platforms may have come across such functionalities (see YouTube captions). This is related to the second criterion, which is the existence of accessibility functionalities in social media platforms. As mentioned in the previous pages, an Alt text option is offered in all programmes and platforms, and (automatic) captioning functionalities are embedded in several of them too. The last criterion was the coverage of diverse groups of user needs. Alt text covers mainly the lack of visual access, captioning the lack of acoustic access, while plain language addresses cognitive load. Therefore, the selected accessibility strategies were alternative text, plain language, and closed captioning.

The infographics were designed trying to achieve the following goals:

1. Create a high impact introduction to basic audiovisual accessibility for those new to the subject.
2. Create examples in which those who do not normally encounter barriers when dealing with audiovisual content, experience them first-hand, along with the associated frustration, eliciting emotions and connecting them to the lessons being learned in order to enhance learning and later recall.

3. Experiment with methods to create accessible infographics.

The structure of the infographics is as follows: First, users are presented with an introductory infographic on web content and accessibility. This is structured along the 'Stop, Think, Act' principle in social skills training (McClelland & Tominey, 2015). This three part social problem solving method builds perceptual skills and self-control (STOP), cognitive problem solving skills (THINK) and behavioural skills (ACT). This structure urges users to stop before posting content, think of the degree of accessibility in their content and act to make it more accessible. The 'ACT' part briefly presents the three accessibility tools. The infographic can be seen in Figure 1: Introductory infographic on Web content &

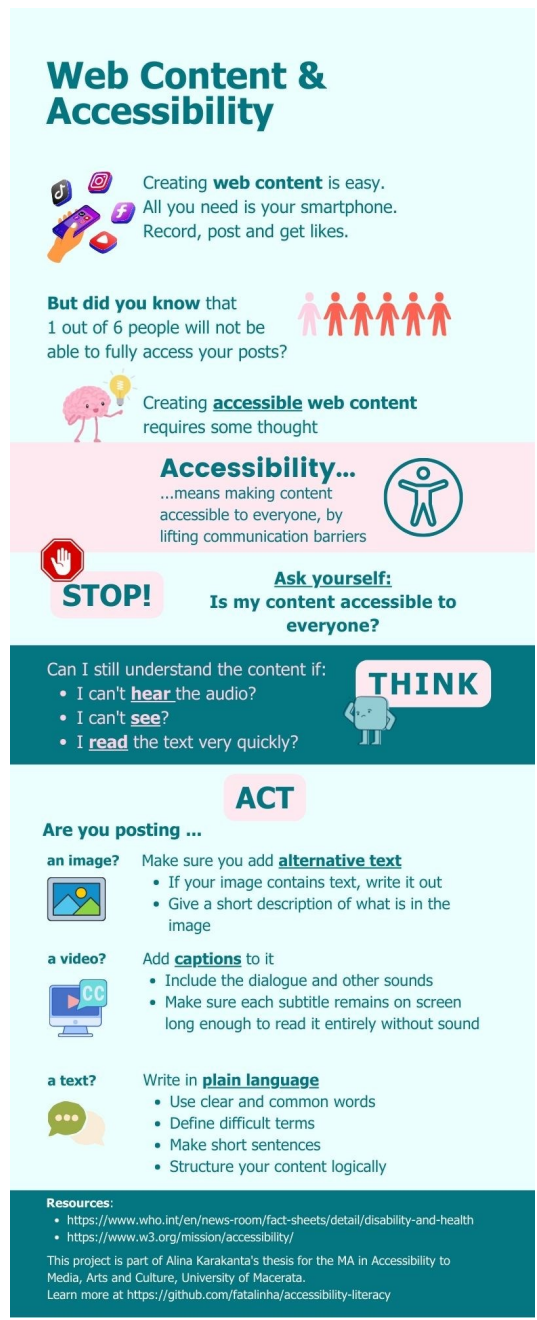

*Figure 1: Introductory infographic on Web content & Accessibility*

The introductory infographic is followed by three infographics, each one on a specific tool: alternative text, plain language, and closed captioning. The infographics have a parallel structure to facilitate learning and navigation. Each infographic is split into four parts. The first part *'What is …'* introduces the notion and presents an example with a barrier. For example, in the Alt text infographic, the text in the image is blurred and the users are asked whether they would be able to understand the meme without having alt text. This allows users to experience the barrier first-hand and think of the effect of their actions on the accessibility of their content. The second part *'Why?'*

follows up from the example by reminding the user of possible scenarios of frustration and explains in which ways each topic can help resolve the barrier that causes the frustration. The third part *'Who it helps'* presents groups of people who may be benefited from the accessibility best practice. These groups are not limited to groups of people with disabilities. For example, Alt text can benefit people who are using their phones in low-data mode, plain language those who are simply tired, while captions are useful for video indexing and information retrieval. This broad range of groups helps the participants to view user diversity as the norm instead of focusing on a specific disability. The last section *'How to'* contains simple instructions on how to implement the accessibility best practices. For instance, plain language contains a list of recommendations with examples, such as *Use active voice, e.g. "we did" instead of "it was done"*. After this section, a button encourages users to take the quiz and test the knowledge they obtained from the infographics. At the bottom, references are included for further reading/watching and information about the project.

The infographics were created using Canva[16]. Canva is an online design and visual communication platform which supports people without design skills in creating content for different purposes (marketing, business, education, etc.). It is particularly suited for creating educational resources by providing templates for tutorials, courses, presentations, posters etc.

A lot of experimentation went into making the infographics accessible while not reducing their pleasantness. The first challenge was the selection of document format. The format needed to support adequately accessibility features but also be easily shared and embedded (e.g. in the survey). Graphics formats, such as .jpg or .png, which are best for sharing and supporting complex images were excluded for the lack of accessibility support. Similarly, the .svg format, even though good for web design and animations, does not support links, which would not allow embedding the quiz link inside the infographic. Portable Document Format (.pdf) was selected, which has good support for accessibility features. The content was structured in a way that it can be properly detected and interpreted by screen readers. Each content unit (header, sentence, or paragraph) was put in separate text box. Interactive features, which may be pleasant and capturing for learners, had to be avoided. The accessibility features are detailed below.

- Colour contrast: The colours of the infographics were selected to have high contrast, with a colour ratio of at least 15. In addition, the combination and colour difference were sufficient for the text to be read by people with colour blindness. For example, white was combined with dark shades of blue or green. The contrast was checked using the Contrast

---

[16] https://www.canva.com/

Checker tool[17], which checks compliance with the Web Content Accessibility Guidelines (WCAG).

- Font: The font used was PT Sans. Being a sans serif font type, which means it does not contain decorative elements, PT Sans allows distinguishing among lowercase "i", lowercase "l", uppercase "L", and numeral "1", as well as between 'mirrored' typography elements, such as "b" and "d" or "q" and "p", which are difficult to distinguish by people with dyslexia and reading difficulties. As for font size, a minimum size of 20px was used for the infographic text and 16px for the references and project information. It should be noted that 16px is the minimum WCAG-conforming font size. Headers were made easily identifiable with a larger font size and bold. Line spacing was set to 1.5.
- Alt text: A description of the information inside the images has been provided when necessary. When the images have a function, alt text contains that, e.g. 'Take quiz'. Images have been marked as decorative when they do not add information to the content.
- Plain language: The principles of plain language were applied. Ambiguous words were avoided and difficult or technical terminology was explained where needed. For instance, instead of using the term 'visual item', the phrase 'image, chart or graph' was preferred. The sentences were kept short and the user was addressed directly through ego-targeting ('Have **you** ever felt frustrated…?'). The content was structured in blocks and bullet lists. In addition, the parallel structure of the three infographics facilitates the understanding of information, since the user becomes familiar with the structure.

The infographics can be seen in Appendix A as image files. For the accessible pdf version, see the data repository.

### Quizzes

Quizzes serve for users to check their understanding of the training on accessibility tools. They iterate the knowledge received from the infographics and contain practical scenarios where users need to apply the tools. The quizzes contain 5-6 multiple choice questions. The first 2-3 questions are related to the content presented in the infographics, checking the attention and understanding of the notions presented. For example, questions include selecting the correct definition or the groups of users who benefit from the tool. The last 3-4 questions are practical examples of how the tool can be used. For plain language, respondents need to select the plain language version between two options, for alternative text, the most suitable alt text for an image and for captions, the most suitable caption given a screenshot from a video. The quizzes are created using Google forms. To apply principles of game-based learning, each question has a specific number of points. The options for each question appear in a randomised order. Respondents can see the correct/right answers after completing the quiz as a way to receive feedback on their understanding.

---
[17] https://contrastchecker.com/

### 3.2. Survey

A survey was designed to collect data for investigating the research questions posed in the introduction. The survey was structured along the following lines:

- Problem: Low usage of accessibility tools among young content creators.
- Program Objective: To expose young creators to accessibility tools and encourage them to include accessibility when creating and sharing web content.
- Indicator: Percent of respondents who are willing to include accessibility in content creation.

The survey was run as a questionnaire, structured in four sections and an introduction. In the introduction, the aim of the study was explained and participants needed to explicitly state their consent. In the first section, demographics about the participants were collected, such as age, gender, country of residence, highest education completed, fields of study, whether they had received professional training on accessibility, as well as questions related to their daily use of social media, frequency of sharing content online and types of content. The second section includes multiple choice questions to establish the current level of accessibility literacy of the participants and their perceptions around accessibility. It registers familiarity with the concept of web accessibility and accessibility guidelines, the consideration of people with disabilities when they create content, whether they believe the content they create is accessible, and whether they have ever experienced barriers in accessing web content and under which circumstances.

The participants are presented with the training material on accessibility in the third part of the survey. The infographics are shown one after the other in pdf format to preserve accessibility features, and participants are asked to respond to 3 quiz questions after each infographic (except for the introductory one), which checks their understanding of the accessibility tool and learning.

The last part aims to assess the usefulness of the training material, the participants' engagement and whether there is a shift in their perceptions on accessibility as a result of the training. It contains 5-point Likert-scale questions (strongly agree-strongly disagree) and two open-ended questions. Some of the Likert-scale questions correspond to questions presented in the second part of the study, to obtain the effect of a pre-test/post-test. A pre-test/post-test is used to observe whether a desired change occurred as a result of the training. In a pre-test/post-test, the same instrument is administered both prior to an activity and after. Answers are compared to identify and report any changes. These questions are *The web content I have been creating so far is accessible to everyone*, and *I believe accessibility is valuable only for people with disabilities*. To avoid confirmatory bias, some question reversals were included. The open-ended questions ask the participants to explain why they would make the content they create accessible or why not and in which ways. This allows to collect more fine-grained and interpretable information on the willingness of the participants to use accessibility tools when creating content.

The survey was designed on the Qualtrics XM experience management platform[18] and is accessible according to the WCAG guidelines. It was piloted with 3 persons who provided feedback on the questions and study presentation. Moreover, testing was performed on different browsers and devices (computer, mobile) to ensure the survey was functional. One of the challenges was to preserve the accessible character of the infographics in the survey. The infographics had to be added as pdf and not image format, which would make the infographic inaccessible. However, embedding pdfs in the Qualtrics survey was not straightforward and the dimensions may not have fitted perfectly in all screen sizes. In addition, one tester reported that they were not able to complete the survey the first time because the quiz opened in the survey window, and this created confusion on how to return to the survey. Since the way a link is opened from a pdf depends on the participant's browser, they only way to ensure that the participants do not close the survey window was to remove the quizzes from the pdf and include quiz questions in the survey. This comes with the drawback that the game-based character of the quiz is lost, since participants do not receive points and feedback on their responses.

The information source for this survey is young persons who create and share content on the web, especially on social media. For this reason, sampling included participants below 35 years of age, who are web content creators. The data collection was performed in autumn 2023 (10 September-10 October). The online survey was distributed through the author's Twitter and LinkedIn account, on the Instagram account of the translation group at Leiden university, and per email at the students of the University of Macerata. The survey is presented in Appendix B.

---

[18] https://www.qualtrics.com

## 4. Results

This section presents the results of the survey. First, details about the participants are presented, followed by the responses on current accessibility literacy. Then, we show the accuracy rate of the quiz responses related to the infographics, and lastly, the perceptions of the participants on the usefulness of the training material and their willingness to use accessibility tools when creating UGC.

### 4.1. Participant demographics

The survey received in total 38 valid and complete responses (67% response rate). The majority of the participants are between 25-34 years old (50%, 12 p.), followed by the group of 18-24 years old (32%, 19 p.) and 7 participants above 35 years old (18%). In terms of gender, 26 were female (68%), 10 male (26%), while 2 participants (5%) preferred not to mention their gender. The geographic distribution of the participants is mainly Europe-centred: 20 participants currently live in Italy, 9 in Greece, 2 in the Netherlands, 2 in Germany, and one participant was reported for the UK, US, Spain, Switzerland and Croatia. The education level of the participants ranged widely: 5 participants (13%) have completed the high school, 10 a university bachelor's degree (26%), 12 a postgraduate or graduate professional degree (32%) and 11 a PhD or similar qualification (29%). Most participants obtained university degrees in fields related to accessibility: 6 participants have a degree in accessibility, 16 in language translation and/or interpreting, 5 in visual or performing arts, 9 in computer science and 1 in design (architectural, industrial, product, web, graphic, etc.). Participants also received degrees in economics, electrical engineering, public health, linguistics and mathematics.

In terms of social media usage, most participants spend 1-2 hours a day on social media (45%, 17 p.), followed by 3-4 hours a day (29%, 11), and less than 30 minutes (26%, 10). As for active posting, 18 participants (47%) post several times a month on social media, with 8 posting more frequently (7 several times a week [18%] and 1 several times a day [3%]), and 7 posting several times a year (18%), while 5 mentioned they never post on social media (13%). The most common type of content shared online by participants is images, pictures and videos (30 responses), followed by texts (15) and videos (13). This confirms that audiovisual content is a very popular type of content shared on social media.

### 4.2. Current accessibility literacy:

To the question *How well do you know the concept of 'web accessibility'?* most participants reported some basic understanding of the concept of web accessibility (16 participants responded *Slightly well* [43%], 9 *Moderately well* [24%], 7 *Very well* [19%]), while 5 reported they did not know the concept at all (14%).

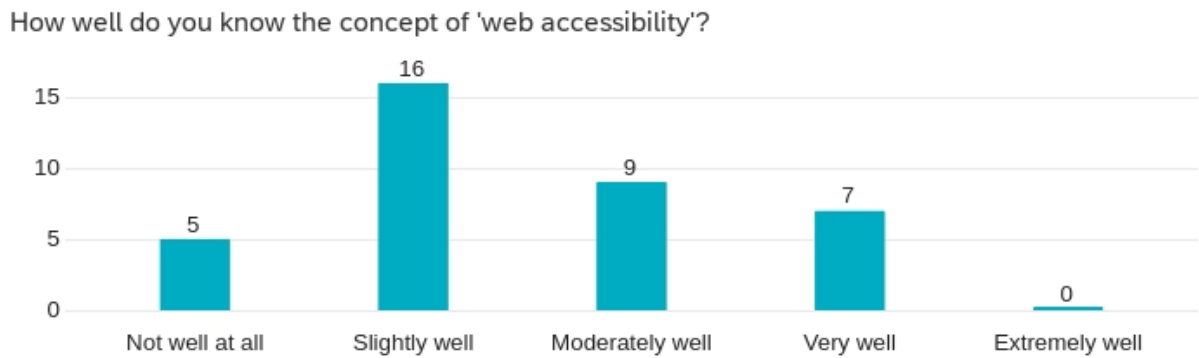

Despite some familiarity with web accessibility, most participants (18 – 49%) reported they were not familiar with any accessibility guidelines (WCAG, ATAG, etc), or were only slightly familiar with them (10 – 27%). This suggests that most participants did not possess the necessary knowledge to implement accessibility tools for their content.

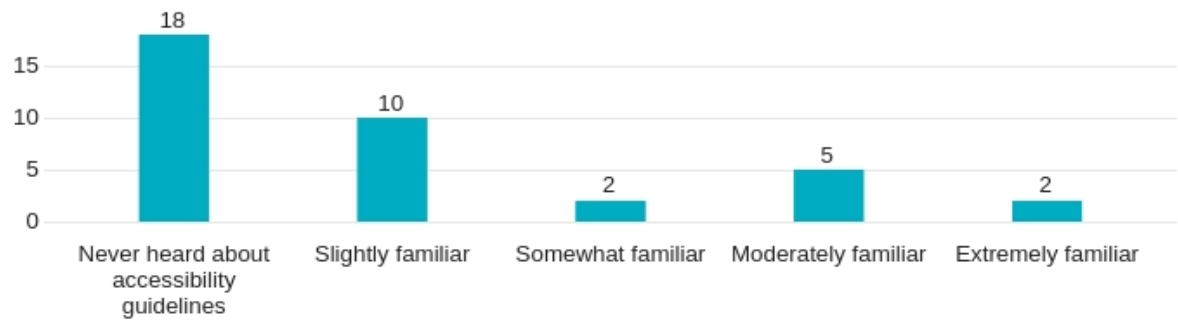

This is corroborated by the following question *Do you think that the content you create and share is accessible by everyone?*, where most answers range between *probably yes* (12 – 32%) and *probably not* (10 – 27%). There is high uncertainty among the participants on whether their content is accessible or not.

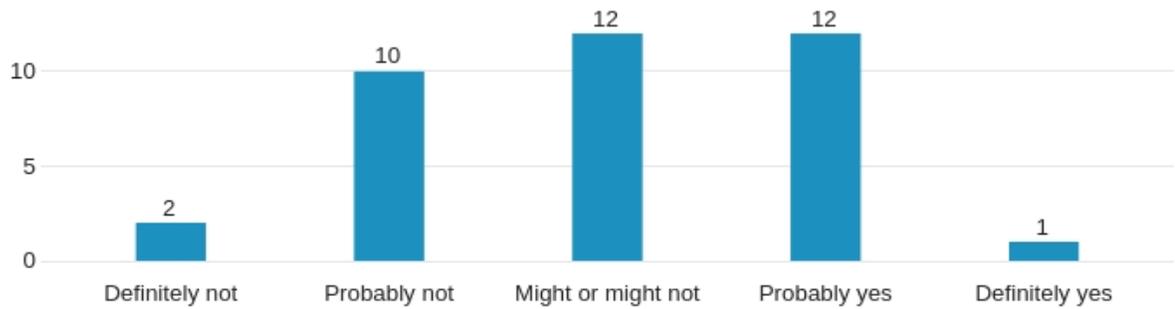

To the question *How often do you think of people with disabilities as users of the content you create and share on the web?*, the most frequent answer was *Sometimes* (17 – 46%), while 7 participants (19%) reported they never think of people with disabilities as users of their content.

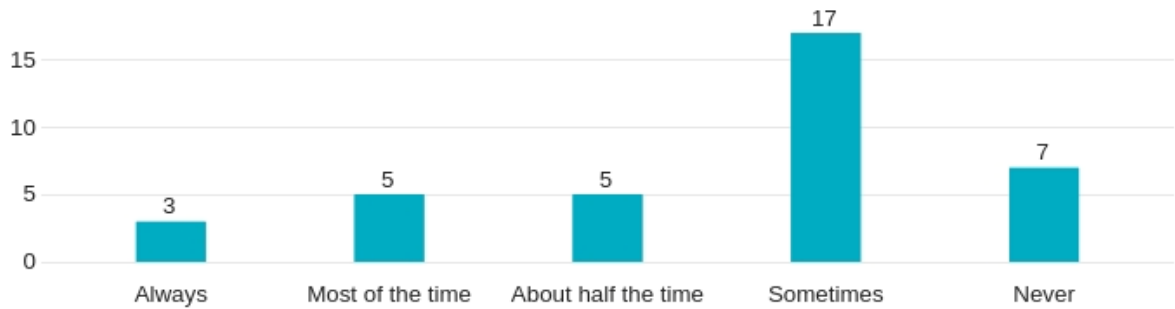

Still, the participants recognised the need to consider accessibility when creating and sharing web content, with 29 participants (88%) saying accessibility should always be considered. Three participants thought that accessibility should be considered only when it is possible given time and cost constraints. The answers to the last two questions suggest that, even though participants have limited awareness about the fact that the way they create content may pose barriers to different groups of people, including people with disabilities, they consider accessibility an important matter in content creation.

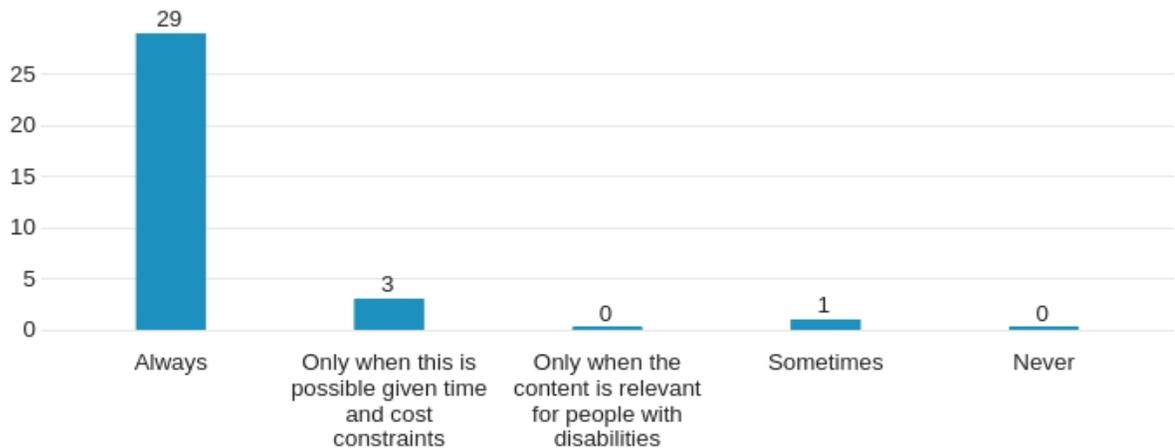

When asked how often they experience barriers that do not allow them to access some web content, the majority of participants did not (16 – 43%) or only sometimes (19 – 51%) experienced barriers. The answers *Most of the time* and *About half the time* received one answer each. The circumstances under which they experienced barriers varied. Two participants mentioned they have a disability. These participants were the ones answering *Most of* and *About half the time* in the previous question, showing that people with disabilities are the ones more often experiencing barriers. Other reasons were having a temporary injury or illness (1), not being able to hear because of noisy environment/low volume (13), not being able to see or read the content because of a broken/faulty screen or environmental factors (e.g. dark room) (7), while the most common barrier is the linguistic barrier when participants did not understand the language of the content (20). Participants also mentioned slow internet connection, websites or applications not respecting user privacy or choice of device or operating system, paywalls, small subtitles and pacy dialogue as additional barriers. One interesting fact is that participants who stated they never experience barriers still selected the linguistic and/or the auditory barrier when prompted by the question. This suggests that participants may not even be aware of the factors that can constitute a barrier.

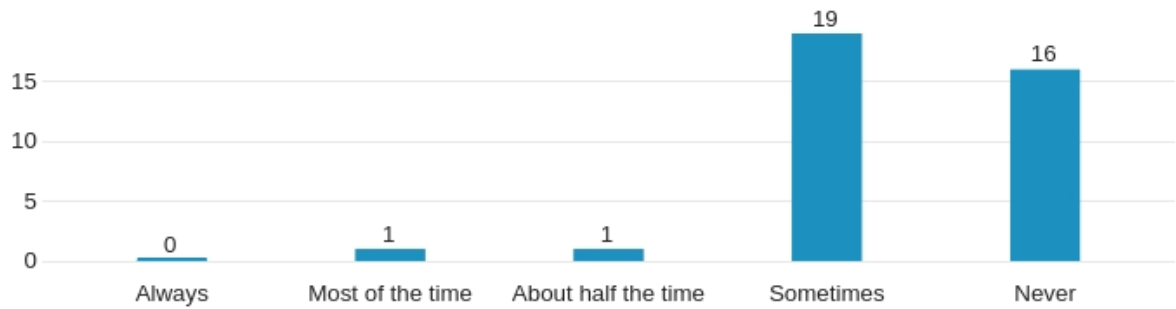

### 4.3. Accuracy of quiz questions

The percentage of participants who responded correctly to the quiz questions for each accessibility tool is shown in Figure 2. We observe that the questions related to alternative text were the most difficult to answer, while for plain text the easiest. In addition, most incorrect answers come from the questions related to the content of the infographics (Q1 and Q2). These questions asked the participants to answer what each accessibility tool is and who it is useful for. The questions where participants were asked to select the correct variant of alt text, plain language or caption (Q3) received a higher percentage of correct answers, even 100% in the case of plain language. It is possible that these questions were relatively easier because the number of options was less than the content questions, so the participants had 33-50% chance to select the correct answer. Despite this, the high percentages suggest that participants have obtained some knowledge of accessibility tools and how to implement them when sharing their content.

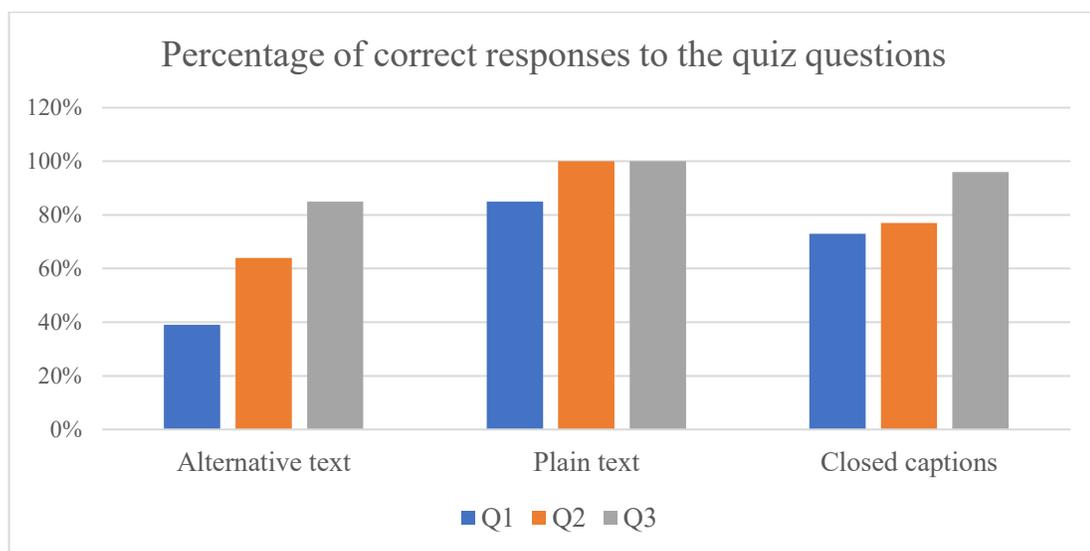

*Figure 2: Accuracy for the quiz questions per accessibility tool.*

### 4.4. Post-task perceptions of accessibility and willingness to implement accessibility tools

Table 1 shows the results of the answers to the Likert scale questions after the participants read the infographics and completed the quiz questions. In this scale, 1 corresponds to 'Strongly agree', 3 to 'Neither agree nor disagree' and 5 to 'Strongly disagree'. The order of the questions has been changed compared to the survey so that similar questions appear together in the table.

| Question | Median | Mean | Min | Max | Variance |
|---|---|---|---|---|---|
| 1) I enjoyed the infographics. | 1.00 | 1.43 | 1.00 | 3.00 | 0.32 |
| 2) I found the information in the infographics clear. | 1.00 | 1.18 | 1.00 | 2.00 | 0.15 |
| 3) I could answer the questions related to the infographics. | 1.50 | 1.50 | 1.00 | 2.00 | 0.25 |
| 4) The infographics helped me develop some knowledge necessary to make my web content more accessible. | 2.00 | 1.78 | 1.00 | 4.00 | 0.77 |
| 5) As a result of the infographics, I am motivated to learn more about accessibility. | 2.00 | 2.07 | 1.00 | 4.00 | 0.64 |
| 6) I discovered faults in what I previously believed as the best way to create and share web material. | 2.00 | 2.43 | 1.00 | 5.00 | 1.46 |
| 7) The web content I have been creating so far is accessible to everyone. | 3.00 | 3.21 | 1.00 | 5.00 | 1.67 |
| 8) The infographics have challenged some of my previous ideas around accessibility. | 2.00 | 2.61 | 1.00 | 5.00 | 1.31 |
| 9) As a result of the infographics, I have changed the way I think about accessibility. | 2.00 | 2.19 | 1.00 | 4.00 | 0.54 |
| 10) I believe accessibility is valuable only for people with disabilities. | 5.00 | 4.70 | 4.00 | 5.00 | 0.21 |
| 11) In the future, I will think more often of people with disabilities as users of my content. | 2.00 | 1.67 | 1.00 | 4.00 | 0.59 |

| 12) It takes too much time and effort to make my content accessible. | 4.00 | 3.75 | 1.00 | 5.00 | 1.33 |
| --- | --- | --- | --- | --- | --- |
| 13) I am willing to apply accessibility tools when creating web content. | 2.00 | 1.63 | 1.00 | 3.00 | 0.38 |
| 14) Making the content I share online accessible will increase its popularity. | 2.00 | 2.30 | 1.00 | 4.00 | 0.80 |

*Table 1: Median, mean, minimum and maximum response values and variance for each of the Likert scale questions. Scale: 1.00=Strongly agree, 2.00=Somewhat agree, 3.00=neither agree nor disagree, 4.00=Somewhat disagree, 5.00=Strongly disagree.*

Questions 1-5 test the usefulness of infographics and quizzes as training material to educate content creators on accessibility issues. The participants strongly agreed with the statements that the infographics were enjoyable and the information contained in them clear, and they were able to answer the quiz questions related to the infographics. They also mostly agreed with the statement that the infographics were helpful for developing knowledge that will allow them to make their content more accessible. In addition, participants found the infographics motivating for them to learn more about accessibility.

Questions 6-11 explore whether there is any change in the participants' perceptions on accessibility after interacting with the training material. Participants somewhat agreed that the infographics have challenged some of their previous ideas around accessibility and that they have changed the way they think about accessibility as a result of the infographics. Moreover, most participants discovered faults in what they previously believed as the best way to create and share web material. Concerning the accessibility of the web material they have been creating, the median answer is 'Neither agree nor disagree', and there is a large variance in responses of the participants (1.67). This is due to the fact that some of the participants were already familiar with accessibility and were already using accessibility tools for their content before taking the study. However, the mean value (3.21) suggests that many participants have not been creating accessible content. The same holds for question 11, *In the future, I will think more often of people with disabilities as users of my content*. Participants who were already aware of accessibility issues somewhat disagreed with this statement. However, the participants who agreed were those who did not consider people with disabilities as users of their content in the question of session 2 *How often do you think of people with disabilities as users of the content you create and share on the web?*. Participants strongly disagreed with the reverse question 12, *I believe accessibility is valuable only for people with disabilities*, showing that they recognised the diverse groups of users who can benefit from accessibility tools. These answers show that the training material was useful in changing the perceptions of the participants on accessibility and disability and in increasing their awareness on accessibility issues.

Lastly, questions 12-14 are related to the willingness of participants to implement accessibility tools in order to make their content more accessible. What is notable is that none of the participants disagreed with the statement *I am willing to apply accessibility tools when creating web content*. Moreover, most participants agree that making their content accessible could increase its popularity. When considering the time and effort required to apply accessibility tools, most participants somewhat disagreed with the statement *It takes too much time and effort to make my content accessible*, even though the answers ranged widely. Still, the responses show that participants recognise the value of accessibility tools despite the effort related to applying them and are willing to use these tools for making their content accessible.

The last part of the survey included two open questions where participants are asked i) to explain why they would make their content accessible in the future or why not, and ii) in which ways they will change the way they create and share web content to make it more accessible. The open questions were analysed based on thematic analysis (Braun and Clarke, 2006), which codes answers based on specific topics.

For the first open question *Please explain why you would make your content accessible in the future or why not*, the majority of the participants mentioned they would make their content accessible in the future. Only two participants said that whether they make their content accessible depends on the effort that it takes and the importance of the content they create. When it comes to the reasons why they would make their content accessible, the thematic analysis revealed three main topics. Firstly, **inclusion** was mentioned in 15 statements. Participants mentioned that accessibility is the only way for everyone to be able to engage and interact with social media and web content. They also stated the importance of accessibility for including everyone, removing barriers and improving the user experience. The second most mentioned reason is to **increase the value** of their content (6 statements). This includes making their content more understandable, making the information presented in it clearer, but also gaining and engaging with a larger audience or connecting with more people. The third reason is the **societal value of accessibility** (5 statements). Participants mentioned that inclusion is important for ensuring equal access and thus society 'needs to work on' accessibility. The ethical side came out in one answer, where the participant claimed that 'it is not right' to exclude people with disabilities. In addition, one participant stated that making their content accessible could possibly help raise awareness on accessibility.

Analysing the attitude of the statements, while most participants mentioned the value of inclusion and making content accessible to everyone, nine (9) statements indicated a view of accessibility from the aspect of the medical model of disability. Participants stated that they would make their content accessible in the future to 'help people with disabilities understand the content', 'to make the life of people easier', or because they 'feel bad for people who can not understand many contents on the Web' and they 'don't like to make people feel different from others'. Other statements specifically mentioned people with 'disabilities' or 'impairments' as the ones who benefit from accessibility, even though the training material took a universal approach to accessibility tools and their usage. When looking into the background of the participants of these statements, all of them had completed high school as the highest level of education and had not received any professional training on accessibility, while more societal views of accessibility came from participants with a higher level of education (Master or PhD) and some (professional) training on accessibility. These statements suggest that the medical model is rooted in the people's perceptions on accessibility but education and training can be a significant factor in shaping these perceptions towards a more universal and user-centred approach on accessibility.

For the second open question, *In which ways will you change the way you create and share web content to make it more accessible?*, the accessibility tools discussed in the infographics are the most common ways mentioned. Alt text was the most mentioned accessibility tool, with 14 statements. This is mainly because images are the type of content that most participants create and share, and thus Alt text is relevant for them. Since videos were the second most popular type of content shared among participants, adding closed captions/subtitles/transcripts appeared in 9 statements. One participant said that they would add captions for more official content, since adding them can be more 'tricky'. Plain language was mentioned in 7 statements. While plain language was found to be useful by one participant especially for website (business) content, they were concerned by budget restrictions. Other ways in which the participants would change the way they create content were selecting an appropriate image scale, being mindful of hashtags, avoiding the use of music without explicitly stating the presence of a song and using text-to-speech. Two participants will not make any change, one of them because they only post texts and the other one because they already implement accessibility tools in their content. One participant stated that if their posts become popular, they will make sure that their 'text and photos are appropriate'.

## 5. Discussion

In this section, the research questions posed in this thesis are discussed on the basis of the results obtained from the survey.

**What is the current literacy level on media accessibility among young content creators?**

The results confirm the initial hypothesis of this thesis and suggest that content creators in general have limited literacy on media accessibility. This is demonstrated by the very basic understanding of the notion of web accessibility, a limited familiarity with accessibility guidelines and the uncertainty on whether the material they create is accessible. The exceptions to this rule are participants who have received education or professional training in accessibility. This shows the importance of incorporating accessibility in study curricula, both at school and university level, for increasing the accessibility literacy and awareness of young creators. Moreover, despite recognising the importance of always considering accessibility on the web, study participants only sometimes think of people with disabilities as users of UGC. In addition to this, participants may not even be aware of the factors that can constitute a barrier. This could be a manifestation of a particularlist view of accessibility, where impairments are the only factors seen as barriers. When prompted to consider other factors as barriers, participants without a disability admitted that they experienced linguistic barriers and lack of auditory and/or visual access due to technical or environmental factors. From these findings we posit that the first step to accessibility literacy is to make people aware of the universal nature of accessibility, before providing knowledge and practical tips on implementing accessibility tools.

**Can training material, such as infographics and quizzes, be used to educate young content creators on accessibility?**

The feedback of the survey participants regarding the training material was very positive. Participants found the infographics clear, pleasant and offering useful information on how to make content accessible. In addition, most participants were able to correctly answer the quiz questions related to the accessibility tools. These findings demonstrate that this type of training material can have some usefulness for educating content creators. However, the circumstances under which the material was administered may not have been optimal. The material was created to be disseminated online and studied independently at one's own pace, which is different from the survey environment, where participants may have limited time and patience to dedicate to the survey. Indeed, the main section where participants dropped-out was the section containing the infographics. It is possible that participants did not find accessibility tools relevant when the training was disconnected from the task of creating and sharing content. Including such material as suggestions inside the social media platforms, where creators can directly see the effect of using accessibility tools, could increase the engagement with the training material.

**Is training helpful to improve perceptions of young content creators towards web accessibility?**

Interacting with the training material seems to have helped improve the perceptions of creators towards accessibility. Participants reported that their previous ideas and thoughts on accessibility were challenged as a result of the infographics and that in the future they will think more of people with disabilities as users of their content. Furthermore, they recognised that accessibility is not useful only for people with disabilities, a notion that was strongly stressed in the infographics by laying out the different groups of users who benefit from accessibility. One point to note is that the majority of participants considered accessibility important in the online sphere even before interacting with the training material, but they were unaware of its nature, its relevance for diverse groups and how to implement it. Even though the extent of accessibility topics in the material as well as its depth had to be limited to fit the infographics and the length of the survey, training influenced the participants' perceptions. Still, the responses to the open questions suggest that ideas stemming from the medical model of disability and a particularlist view of accessibility are deeply rooted and more profound attempts are required to shift such perceptions, possibly inside critical learning spaces.

**Are young content creators more willing to create accessible content after receiving training on web accessibility?**

All participants expressed their willingness to create accessible content after receiving the training. They stated that accessible UGC is important for improving the content they create, for avoiding excluding and marginalising persons and for 'doing the right thing'. Reaching and engaging wider audiences was also a motivation for them to consider accessibility. However, some noted that whether they will indeed implement accessibility tools depends on the effort required and the importance of the content. The ways they would make their content accessible depend on the type of content they mostly share, creators sharing images were more interested in Alt text, while those sharing videos stated they would try to include captions. The majority of the participants mentioned they will use at least one accessibility tool of the ones presented in the infographics. All in all, the content creators participating in this survey seem willing to create more accessible content as a result of the training, at least in theory and for the purposes that serve their content. It is left to future studies to measure the duration of the training effect on the habits of content creators.

This study does not come without limitations. Firstly, the sample of the participants to the study may not have been utterly representative of the wide range of content creators and users of social media. The number of valid responses (38) may have been sufficient for running statistical analyses but is still limited. The participants were largely based in Europe, coming from western developed countries, while there were no participants currently living in the south hemisphere. In addition, participants who had received training on accessibility gave different responses than those who had not. Follow-up studies should consider two groups, those with and without training on accessibility, and compare the differences in perceptions between the two groups. Secondly, the survey environment restricted the interaction with the training material to a large extent. The integration of the pdf files in the survey was an elaborate process and their display on mobile devices required scrolling to see the entire content. Because of the risk of closing the survey window before completing it, the quizzes had to be included as normal survey questions, which removed the game-based character of the training material. Participants may have found the survey too long, considering the drop-out rate, even though those who completed the survey did so in average 12 minutes. A different setting, where participants are allowed to freely interact with the training material for as long as they wish, would have been a more suitable learning environment. However, this would not make it possible to distribute the survey online and would require recruiting participants. In turn, this would restrict the geographic range of the participants even more. Despite these limitations, the study has provided useful insights on content creators' attitudes on accessibility and a roadmap for future directions for enhancing accessibility of UGC.

# 6. Conclusions

This research project has been devoted to enhancing the accessibility literacy of young content creators, with the ultimate objective of fostering more inclusive user-generated content. Achieving this goal involved the design of simple and easy-to-use training material aimed at raising awareness among content creators about accessibility tools and equipping them with starting knowledge to create accessible content.

The study addressed four key research questions:

1. What is the existing level of media accessibility literacy among young content creators?
2. Can training materials, such as infographics and quizzes, effectively educate young content creators on accessibility?
3. Does training significantly improve the perceptions of young content creators regarding web accessibility?
4. Are young content creators more inclined to create accessible content after receiving training on web accessibility?

The research effort unfolded in two distinct phases. Initially, training materials were developed encompassing informative infographics and engaging quizzes that focused on basic web accessibility tools. Subsequently, a survey was designed to capture data concerning the current accessibility literacy of young creators and the effectiveness of the training materials in reshaping their perspectives on media accessibility.

In conclusion, the results of this thesis project provide valuable insights into the accessibility literacy of young content creators, both before and after receiving accessibility training. The findings reveal that while many of these creators initially possess limited accessibility knowledge, even simple training materials can have a positive impact on their perceptions of web accessibility. Furthermore, the survey indicates that these creators are generally open to changing their content creation methods to enhance accessibility, with the specific approaches varying based on the type and purpose of their content. Although there were instances of a medical model of disability and a particularist view of accessibility in some responses, it is clear that accessibility is widely recognized as a key factor in promoting inclusivity, enhancing user-generated content, and fostering a fairer society for all.

The findings of this study have provided insights on practices which could help improve creators' perceptions of accessibility and lead to more accessible UGC. These insights emphasize the importance of raising awareness about accessibility in a way that promotes the universalist account, before providing any practical information into the implementation of accessibility tools. The universalist account can broaden the scope of accessibility so that creators can identify themselves and those around them as persons who can benefit from it. We observed that even a short exposure to accessibility material helped shift perceptions, therefore integrating accessibility training into formal educational curricula of different levels and fields could be implemented with small but focused interventions. When it comes to UGC, users may be motivated to make it accessible, but they cannot succeed if accessibility functionalities are not provided by the platforms and media where UGC is created and shared. For this reason, exerting pressure on social media platforms to incorporate effective accessibility features right from the outset is paramount to accessible UGC. Furthermore, platforms could integrate recommendations guiding users on how to make their posts more accessible, such as using prompts like, "Your post contains an image. Have you remembered to add Alt text?". Accessible social media could finally be considered as an opportunity for people to be reminded of the diversity of user needs and consider accessibility at a daily basis.

As screens become windows to distant realities and stories come alive with a symphony of senses, catering for diverse user needs in this audiovisual revolution is essential for forming a fair and diverse landscape of modern communication. Making user-generated content more inclusive and accessible for everyone is a step into that direction.

- Waller, A., Hanson, V. L., & Sloan, D. (2009). Including accessibility within and beyond undergraduate computing courses. In *Proceedings of the 11th international ACM SIGACCESS conference on Computers and Accessibility (Assets '09)* (pp. 155–162). Association for Computing Machinery. https://doi.org/10.1145/1639642.1639670
- Web Content Accessibility Guidelines 3.0. (2023). Retrieved August 15, 2023, from https://www.w3.org/TR/wcag-3.0/
41

# Appendix

## A. Infographics

Introductory infographic

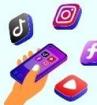

Alternative text

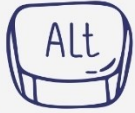
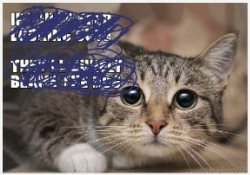
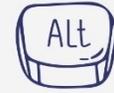
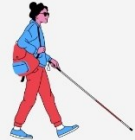
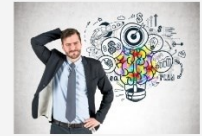
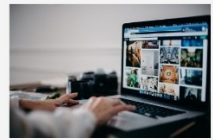
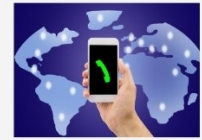

Plain language

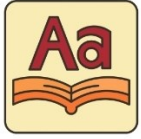 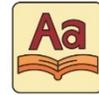 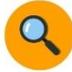 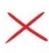 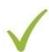 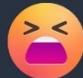 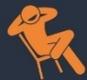 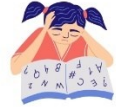 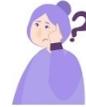 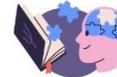 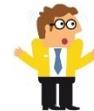 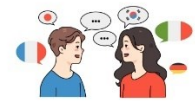 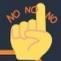

Closed captions

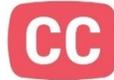
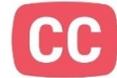

# How to make your content accessible using Closed Captions

## What is closed captions?

**Closed captions** (or Subtitles for the Deaf and hard of hearing) is a text version of the speech and other audio information needed to understand content. Captions include:
- the words that are spoken
- who is speaking when it is not evident
- sounds like music, laughter, and noises

**Can you make sense of the following scenes without looking at the captions?**

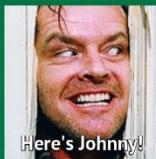
Here's Johnny!

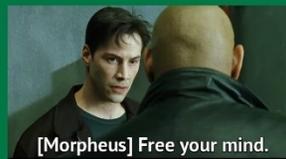
[Morpheus] Free your mind.

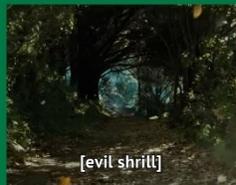
[evil shrill]

## Why?

Have you ever felt frustrated when watching a video without hearing the sound, e.g. in a very noisy environment?

**With captions, audio content is:**
- **accessible** to people with hearing impairments
- easier to **understand**
- easier to **skim, explore and search**

Captions make access to audio information easier for everyone.

## Who it helps?
**People who....**

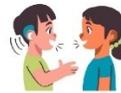
are Deaf or hard of hearing

watch videos in silent mode, e.g. in the metro

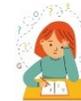
have cognitive and learning disabilities and need to see and hear the content

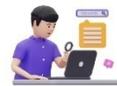
want to easily search for information inside videos

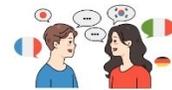
are learning a language or are non-native speakers

## How to
- You can use a subtitle editor to caption your videos. Often such tools are built into social media platforms.
- Captions are one or two lines. If possible, place each sentence on a different line.
- It is best to keep them under 32 characters per line.
- Captions have to be in sync with the speech.
- Captions need to stay on screen long enough for viewers to be able to read them comfortably.
- Voice recognition software can help create automatic captions for some types of videos, but their quality may be poor. Plan sufficient time for correcting them!

**Test your knowledge! Solve the Captions Quiz!**

**References:**
- https://www.w3.org/WAI/perspective-videos/captions/

This project is part of Alina Karakanta's thesis for the MA in Accessibility to Media, Arts and Culture, University of Macerata. Learn more at https://github.com/fatalinha/accessibility-literacy

## B. Survey questionnaire

# Accessibility literacy

**Start of Block: Informed Consent**

Q1
Welcome to the research study on Accessibility Literacy!

This study is about accessibility literacy. It aims to understand whether people who create and share web content are aware of accessibility tools and how awareness of accessibility can be improved. Web content covers posts with videos, images or simple text which can be shared on social media and other platforms, such as websites, blogs, Facebook, YouTube, WhatsApp, Instagram, LinkedIn, TikTok, Twitter, and many more.

For this study, you will be shown four infographics on accessibility topics. Then, you will be asked to answer a few questions about them. The study should take around 20 minutes to complete. Your responses will be kept completely confidential. Your participation in the study is voluntary. You have the right to withdraw at any point during the study. If you have any questions about this study, please contact Alina Karakanta at alinakarakanta@gmail.com.

By clicking the button below, you acknowledge:
Your participation in the study is voluntary.   You are aware that you may choose to terminate your participation at any time for any reason.

○ I consent, begin the study  (1)

○ I do not consent, I do not wish to participate  (2)

*Skip To: End of Survey If Welcome to the research study on Accessibility Literacy!     This study is about accessibility li... = I do not consent, I do not wish to participate*

Q9 How old are you?

○ Under 18  (1)

○ 18-24 years old  (2)

○ 25-34 years old  (3)

○ Above 35 years old  (4)

Q10 How do you describe yourself?

- ○ Male  (1)
- ○ Female  (2)
- ○ Non-binary / third gender  (3)
- ○ Prefer to self-describe  (4) ________________________________________________
- ○ Prefer not to say  (5)

---

Q38 Which country do you currently live in?

________________________________________________

Q11 What is the highest level of education you have completed?

- ○ Less than high school  (1)
- ○ High school  (2)
- ○ University Bachelors Degree  (3)
- ○ Postgraduate or graduate professional degree (MA, MS, MBA, JD, MD, DDS etc.)  (4)
- ○ PhD or similar  (5)
- ○ Prefer not to say  (6)

---

---

*Display This Question:*

*If What is the highest level of education you have completed? = University Bachelors Degree*

*Or What is the highest level of education you have completed? = Postgraduate or graduate professional degree (MA, MS, MBA, JD, MD, DDS etc.)*

*Or What is the highest level of education you have completed? = PhD or similar*

---

Q6 What is the field of study of your university degree? If you have multiple university degrees, please select all the relevant fields.

- ☐ Accessibility  (1)
- ☐ Language Translation and/or Interpreting  (2)
- ☐ Visual or performing acts  (3)
- ☐ Cultural heritage  (7)
- ☐ Computer science, software engineering  (4)
- ☐ Design (architectural, industrial, product, web, graphic, etc.)  (5)
- ☐ Other (please specify,  add all the fields needed)  (8) ________________________________________________

Q40 Have you received any professional training on accessibility?

- ○ Yes  (1)
- ○ No  (2)

Q7 How much time do you spend on social media each day?
 (Social media include websites and apps, e.g. Facebook, YouTube, WhatsApp, Instagram, TikTok, Twitter, etc.)

- ○ Less than 30 minutes  (1)
- ○ 1-2 hours  (2)
- ○ 3-4 hours  (3)
- ○ More than 4 hours  (4)

Q8 How often do you share content on social media and other online platforms?

- ○ Several times a day  (1)
- ○ Several times a week  (2)
- ○ Several times a month  (3)
- ○ Several times a year  (4)
- ○ I never post on social media  (5)

Q31 What types of content do you share on the web? (check all that apply)

- ☐ Videos  (1)
- ☐ Images, pictures, photos  (2)
- ☐ Simple text  (3)
- ☐ Other (please specify)  (4) ________________________________________________
- ☐ I never share content on the web  (6)

**End of Block: Personal data**

**Start of Block: Current level of accessibility literacy**

Q32 How well do you know the concept of 'web accessibility'?

- ○ Not well at all  (1)
- ○ Slightly well  (2)
- ○ Moderately well  (3)
- ○ Very well  (4)
- ○ Extremely well  (5)

Q31 What types of content do you share on the web? (check all that apply)

☐ Videos (1)

☐ Images, pictures, photos (2)

☐ Simple text (3)

☐ Other (please specify) (4) ________________________________________________

☐ I never share content on the web (6)

**End of Block: Personal data**

**Start of Block: Current level of accessibility literacy**

Q32 How well do you know the concept of 'web accessibility'?

○ Not well at all (1)

○ Slightly well (2)

○ Moderately well (3)

○ Very well (4)

○ Extremely well (5)

Q32 How well do you know the concept of 'web accessibility'?

○ Not well at all  (1)

○ Slightly well  (2)

○ Moderately well  (3)

○ Very well  (4)

○ Extremely well  (5)

Q34 How often do you experience barriers that do not allow you to access some web content?

○ Always  (1)

○ Most of the time  (2)

○ About half the time  (3)

○ Sometimes  (4)

○ Never  (5)

---

Q35 If you have experienced barriers, under what circumstances? (Check all that apply)

☐ I have a disability  (1)

☐ I had a temporary injury or illness  (2)

☐ I couldn't hear because of a noisy environment/low volume  (3)

☐ I couldn't see or read the content because of a broken/faulty screen or environmental factors (e.g. dark room)  (4)

☐ I didn't understand the language of the content  (5)

☐ Other  (6) ____________________________________________

**End of Block: Current level of accessibility literacy**

**Start of Block: Accessibility training**

Q19 In this part of the survey, you will be shown four infographics. Please go through them one by one. You can take as much time as you need for each infographic. The first infographic will provide you with general information about web accessibility and basic tools. Then, infographics 2, 3 and 4 will provide you with specific information about a specific tool. After infographics 2, 3 and 4, you will be asked a series of questions where you can check your understanding of the topics.

Q21 View the infographic about web accessibility. When you are finished, click on the >> button at the end of the page to continue the survey. There are no questions after this infographic.

[Web Content & Accessibility]

Q24 View the infographic about Alternative text. When you are finished, click on the >> button to proceed with answering the questions.

[Alternative text infographic]

Q27 Alternative text is useful for:

○ people who have difficulty perceiving visual content can use assistive technology to read text aloud, present it visually, or convert it to braille.  (1)

○ people who have weak internet connection or are using data-roaming.  (2)

○ bots trying to index your image and web page.  (3)

○ all of the above.  (5)

○ none of the above.  (4)

---

Q28 Choose the most suitable alt text for the image below:

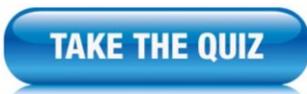

○ Image of a button saying: "Take the quiz"  (1)

○ https://myquiz.com/alt-text  (2)

○ Take the quiz  (3)

---

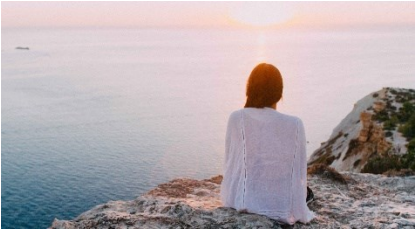

Q29 Imagine you were posting this picture on Instagram. Choose the most suitable alt text for the image below:

○ Woman sitting on cliff looking over the ocean at sunset  (1)

○ COOLEST HOLIDAYS EVER! :) #myconos #summer  (2)

○ Image of the sunset and sea  (3)

○ No alt text necessary (" ")  (4)

Page Break

Q25 Click on the text below to view the infographic about Plain language. When you are finished, click on the >> button to proceed with answering the questions. [Plain Language Infographic]

Q32 Plain language is:

○ language that allows users to easily find what they need, understand what they find and use that information.  (1)

○ oversimplified language, so that people with disabilities can understand all the content.  (2)

○ language that can only be used for non-technical texts.  (3)

Q34 Choose the version that is in plain language.

○ For the chocolate cake you will need 200 grams of butter, 150-200 grams of sugar, 3 large eggs, 200 grams of all-purpose flour, 50 grams baking soda.  (1)

○ For the chocolate cake you will need:
200 grams of butter,
150-200 grams of sugar,
3 large eggs,

200 grams of all-purpose flour,
50 grams baking soda   (2)

Q35 Choose the version that is in plain language.

○ Protect your community from natural disasters!  (1)

○ Mitigation is the cornerstone of emergency management to lessen the impact disasters have on people's lives and property.  (2)

Page Break

Q26 Click on the text below to view the infographic about Closed captions. When you are finished, click on the >> button to proceed with answering the questions.
[Closed captions Infographic]

Q36 Closed captions include: (check all that apply)

☐ a transcription of the words that are spoken  (1)

☐ words describing sounds that are important to understand the plot  (2)

☐ comments from the subtitler describing difficult cultural terms  (3)

☐ the name of the person who is speaking when this person is not visible  (4)

☐ none of the above  (5)

Q37 Closed captions are useful:

○ only for people with hearing impairments, e.g. the Deaf and older people who are losing their hearing  (1)

○ for any person who does not have permanent or temporary access to the audio or who needs access to the text version of the spoken information  (2)

○ for people who are watching videos in silent mode or in noisy environments  (3)

○ for video platforms  (4)

Q38 Select the most suitable caption for the scene below, taken from a video:

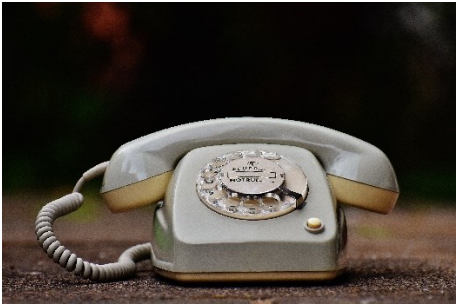

○ [phone rings]  (1)

○ "Ring, ring!"  (2)

○ No caption needed  (3)

Q36 Please select the option that best describes your opinion on each topic:

| | Strongly agree (1) | Somewhat agree (2) | Neither agree nor disagree (3) | Somewhat disagree (4) | Strongly disagree (5) |
|---|---|---|---|---|---|
| I enjoyed the infographics. (1) | ○ | ○ | ○ | ○ | ○ |
| I found the information in the infographics clear. (2) | ○ | ○ | ○ | ○ | ○ |
| I could answer the questions related to the infographics. (3) | ○ | ○ | ○ | ○ | ○ |
| I discovered faults in what I previously believed as the best way to create and share web material. (11) | ○ | ○ | ○ | ○ | ○ |
| As a result of the infographics, I am motivated to learn more about accessibility. (4) | ○ | ○ | ○ | ○ | ○ |
| The web content I have been creating so far is accessible to everyone. (5) | ○ | ○ | ○ | ○ | ○ |
| The infographics helped me develop some knowledge necessary to make my web content more accessible. (6) | ○ | ○ | ○ | ○ | ○ |
| The infographics have challenged some of my previous ideas around accessibility. (7) | ○ | ○ | ○ | ○ | ○ |
| I believe accessibility is valuable only for people with disabilities. (9) | ○ | ○ | ○ | ○ | ○ |
| It takes too much time and effort to make my content accessible. (13) | ○ | ○ | ○ | ○ | ○ |
| In the future, I will think more often of people with disabilities as users of my content. (10) | ○ | ○ | ○ | ○ | ○ |
| I am willing to apply accessibility tools when creating web content. (12) | ○ | ○ | ○ | ○ | ○ |
| Making the content I share online accessible will increase its popularity. (14) | ○ | ○ | ○ | ○ | ○ |
| As a result of the infographics, I have changed the way I think about accessibility. (15) | ○ | ○ | ○ | ○ | ○ |

Q37 Please explain why you would make your content accessible in the future or why not.

_______________________________________________

_______________________________________________

Q41 In which ways will you change the way you create and share web content to make it more accessible?

_______________________________________________

*End of Block: Post-task questionnaire*